# Overview of the SIM PlanetQuest Light (SIM-Lite) mission concept.

R. Goullioud[*], J. H. Catanzarite, F. G. Dekens, M. Shao, J. C. Marr IV,
Jet Propulsion Laboratory, California Institute of Technology, 4800 Oak Grove Dr, Pasadena, CA

## ABSTRACT

The Space Interferometry Mission PlanetQuest Light (or SIM-Lite) is a new concept for a space borne astrometric instrument, to be located in a solar Earth-trailing orbit. SIM-Lite utilizes technology developed over the past ten years for the SIM mission. The instrument consists of two Michelson stellar interferometers and a precision telescope. The first interferometer chops between the target star and a set of Reference stars. The second interferometer monitors the attitude of the instrument in the direction of the target star. The telescope monitors the attitude of the instrument in the other two directions.

SIM-Lite will be capable of one micro-arc-second narrow angle astrometry on magnitude 6 or brighter stars, relative to magnitude 9 Reference stars in a two degree field. During the 5 year mission, SIM-Lite would search 65 nearby stars for planets of masses down to one Earth mass, in the Habitable Zone, which have orbit periods of less than 3 years. SIM-Lite will also perform global astrometry on a variety of astrophysics objects, reaching 4.5 micro-arc-seconds absolute position and parallax measurements. As a pointed instrument, SIM-Lite will be capable of achieving 8 micro-arc-second astrometric accuracy on 19th visual magnitude objects and 15 micro-arc-second astrometric accuracy on 20th visual magnitude objects after 100 hours of integration.

This paper will describe the instrument, how it will do its astrometric measurements and the expected performance based on the current technology.

## 1. INTRODUCTION

The Space Interferometry Mission PlanetQuest (SIM PQ) is a space borne instrument [1] that would carry out astrometry to micro-arc second precision on the visible light from a large sample of stars in our galaxy and search for earth-like planets around nearby stars [2]. SIM PQ would carry out its 5 year mission from an Earth-Trailing Solar Orbit. SIM PQ is a key project of NASA's Navigator program and Search for Earthlike planets and life. The SIM PQ instrument is an optical interferometer system with a baseline of 9 meters, and includes two "Guide" interferometers for spacecraft pointing Reference and a "Science" interferometer to perform high accuracy astrometric measurements on target stars. However, due to its high cost, the SIM PQ mission has been postponed indefinitely.

SIM PlanetQuest-Light (or SIM-Lite) is a cost effective alternative concept for SIM PQ. With a smaller 6 meter baseline and a 30 cm Guide telescope instead of a third interferometer, SIM-Lite will still produce a large amount of the original SIM PQ science objectives. The overall cost of SIM-Lite, slightly above half of the SIM PQ cost to complete, makes SIM-Lite an attractive candidate for flying an astrometric mission by the middle of the coming decade.

The primary objective for SIM-Lite (46% of the 5 year mission) is to search 65 nearby stars for exoplanets of masses down to one Earth mass, in the Habitable Zone. Fig. 1 shows the discovery space for planets with SIM-Lite. For planets with orbit between 100 days and 10 years, SIM-Lite would be better than any other technique by a factor of 100 for the selected 65 stars, enabling the search down to one Earth mass planet. GAIA (a European mission currently under development) will be limited to the search of Jupiter size planets.

---

[*] Send correspondence to: Renaud.Goullioud@jpl.nasa.gov, phone: 1 818 354 7908

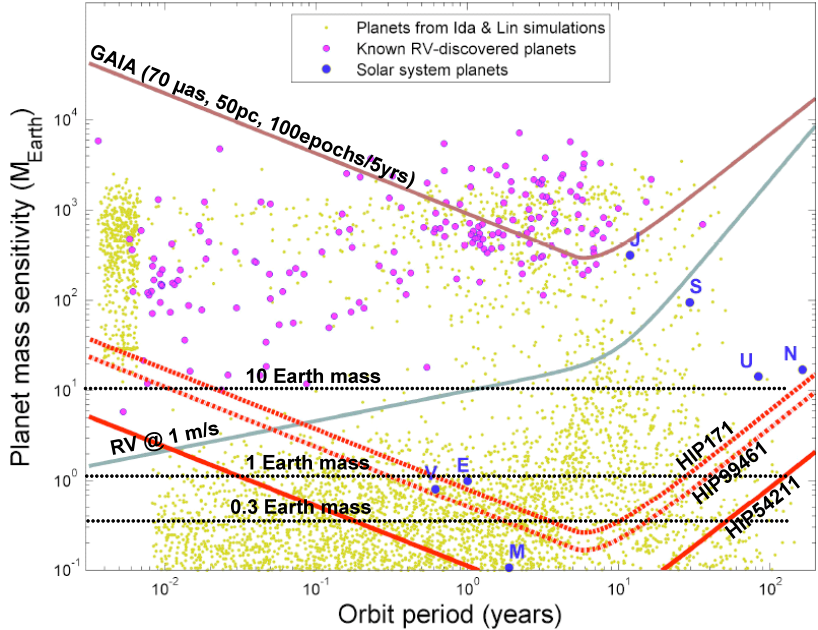

Fig. 1. Discovery space for planets with SIM-Lite. The expected performance for SIM-Lite is shown for 3 of the 60 target stars (Hip. 171, Hip. 99461, Hip. 54211). Planet detection limit for GAIA and for Radial Velocity at 1 meter per second is shown for comparison.

SIM had many technological challenges to address in order to show the mission was technically achievable. These challenges range from nanometer-level control problems to picometer-level sensing problems [3]. Key testbeds and brass-board components have been designed, built, and tested during the technology development phase of SIM [4], resolving all the major technology challenges. Examples of such demonstrations include the System Test-Bed 3 [5], the Micro-Arc-second Metrology testbed [6], the Kite testbed [7], the Thermal-Opto-Mechanical testbed [8] and the Spectral Calibration development Unit [9]. The results from these series of testbeds form the evidence that the technological challenges faced by SIM are achievable. This technology developed for SIM PQ directly applies to the SIM-Lite mission.

## 2. INSTRUMENT

### 2.1. Astrometry with an interferometer

The basic elements of a stellar interferometer are shown in Fig. 2. Light from a distant source is collected at two points and combined using a beam splitter. The interference of the combined wavefronts produces fringes when the internal pathlength difference (or delay) compensates exactly for the external delay. Thus, the angle between the interferometer baseline and the star can be found using the measured internal optical path difference (OPD), according to the relation:

$$\cos\alpha = \frac{\vec{B}\cdot\hat{s}}{B} = \frac{x}{B} \qquad (1)$$

where $x$ is the relative delay (OPD) of the wavefront to one side of the interferometer due to the angle. Thus, the astrometric angle $\alpha$ between the interferometer baseline and the ray from the star can be measured if the length of the baseline B and the internal delay are measured. In a stellar interferometer, the external metrology system measures the distance between two fiducials (each made of common-vertex corner cubes) and the internal metrology measures the optical path difference to the beam combiner from the two fiducials. Finally the starlight fringe detector measures the total optical path difference all the way to the star.

SIM-Lite, like SIM PQ, is a Michelson interferometer operating in the visible spectrum. Light from a star is collected by two 30cm telescopes separated by a 6 meter baseline. From the two collecting telescopes, the light is propagated by a set of optics to the beam combiner where the two optical wavefronts are re-combined, forming interferences. The peak interference fringe is obtained when the propagation path through the two arms of the instrument is identical. The internal metrology sensor measures the internal propagation difference between the two arms (also known as internal delay), from the fiducials on the collecting optics all the way to the re-combining optic, to the single picometer accuracy. When the instrument is tracking the peak interference fringe, the external delay is the complementary of the internal delay. Therefore, the measurement of the internal delay is an excellent estimation of the external propagation delay. Simultaneously, the external metrology sensor determines the length of the baseline, defined by the two fiducials, to similar picometer-level accuracy. Fig. 2 shows that the ratio of the external delay to the baseline length is the cosine of the angle between the baseline and the star.

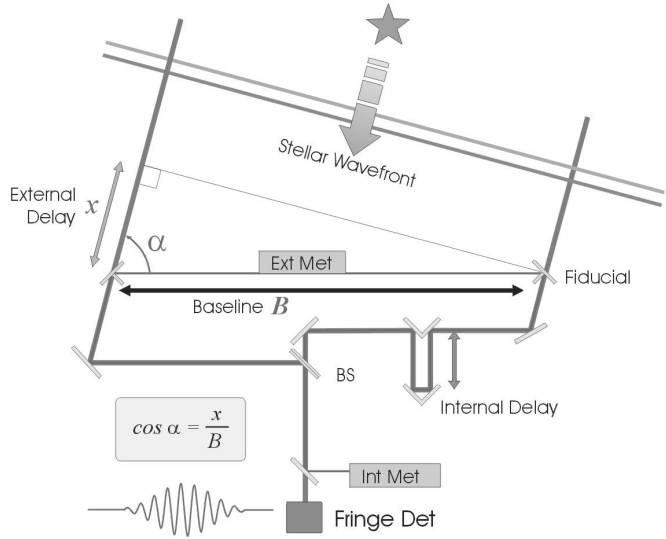

Fig. 2. Astrometry with a stellar interferometer. The starlight fringe contrast is maximum when the internal delay matches the external delay.

SIM-Lite simultaneously employs two stellar interferometers and one Guide telescope to perform astrometry. Fig. 3 shows their relative orientation within the optical configuration. Precision astrometry requires knowledge of the baseline change in orientation to the same order of precision as the astrometric measurement. To achieve this, a second stellar interferometer is required to measure the baseline orientation in the most sensitive direction and a high-precision telescope to measure the baseline orientation change in the other two directions. The second interferometer and the precision telescope acquire and lock on bright "Guide" stars, respectively named "Guide 1" and "Guide 2", keeping track of the uncontrolled rigid-body motions of the instrument. In the meantime, the main interferometer, switches between Science targets, measuring projected angles between them.

The contribution from the Guide 2 telescope to the final astrometric measurement is captured by the Guide 2 scale factor; $\alpha$ is the angular separation from the Science star to the Guide 1 star and $\varphi$ is the angular separation between the Guide 2 star to the Guide 1 interferometric plane:

$$\text{Guide 2 scale factor} = \sin(\alpha)/\sin(\varphi) \qquad (2)$$

To achieve maximum accuracy for narrow angle astrometry, the stars observed with the Science interferometer shall be limited to a 1 degree radius field around the Guide 1 star. The Guide 2 star on the other hand shall be located at 90 degrees away from the Guide 1 interferometric plane. In this case, a 100 micro-arc-seconds error in the measurement of the position of the Guide 2 star produces only a 0.7 micro-arc-second error on the astrometric measurement, while a 1 micro-arc-second error in the measurement of the position of the Guide 1 star still produces a 1 micro-arc-second error on the astrometric measurement. This 140 times relaxation on the required accuracy of the Guide 2 star measurement was the big driver for downgrading SIM PQ's third interferometer to a single telescope on SIM-Lite. The Guide 2 scale factor however increases linearly with the angular separation α from the Science star to the Guide 1 star, limiting the SIM-Lite performance for wide angle astrometry over the full 15 degrees field of regard.

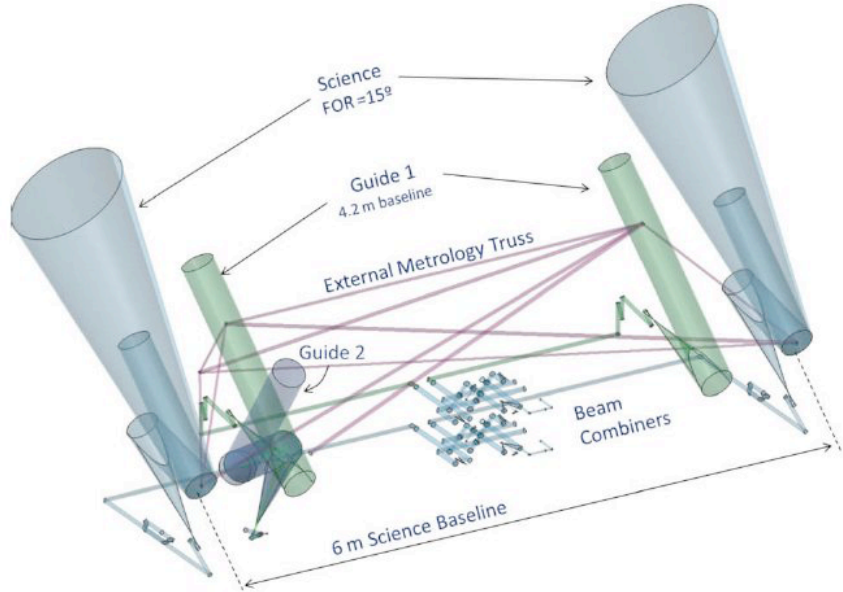

Fig. 3. SIM-Lite optical configuration.

A simple error budget in Fig. 4 shows how the 1.4 micro-acrsecond narrow angle differential measurement accuracy is sub-allocated between the various sensors.

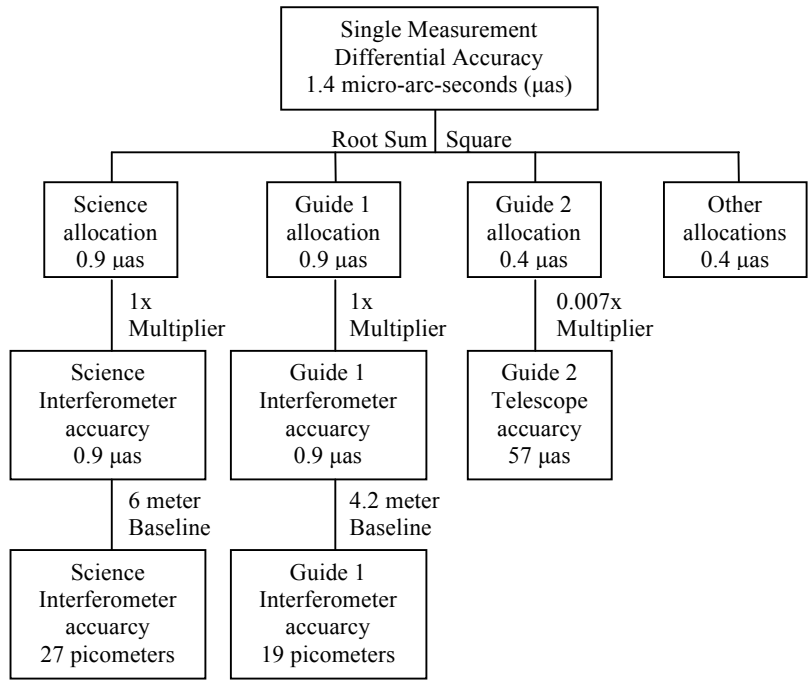

Fig. 4. Simplified SIM-Lite astrometric error budget

## 2.2. Science Interferometer

The Science interferometer consists of two 30cm siderostats separated by a 6 meter baseline. Light reflecting from the two siderostats, is collected by optical confocal beam compressors and propagated by a set of optics to the beam combiner where it is re-combined, forming interferences. The siderostats have an angular range of articulation that enables acquisition of stars in a 15 degree diameter field in order to build the grid described in the "Global astrometric grid" section of the paper. An optical delay line system with 0.8 meter of travel range produces the needed range of internal delay that enables fringe acquisition in that 15 degree diameter field of regard.

## 2.3. Guide 1 Interferometer

The Guide 1 interferometer consists of two fixed 30cm telescopes separated by a 4.2 meter baseline. Light collected by the telescopes propagates through a set of optics to the beam combiner where it is re-combined, forming interferences. The Guide 1 has a very narrow field of regard of only a few arc-seconds, just enough to compensate for errors in the pointing of the entire spacecraft.

## 2.4. Guide 2 Telescope

The Guide 2 telescope consists of a 30cm siderostat and a 30 cm confocal optical beam compressor, similar to the other four telescopes in the Science and Guide 1 interferometers. Light collected by the telescope reflects on a relay mirror and propagates directly to the pointing detector as shown in Fig. 5. The detector arrangement of the metrology system for the Guide 2 Telescope is designed differently in order to measure tip-tilt instead of piston. The metrology light is injected through the relay mirror and monitors the pointing of the siderostat mirror relative to the Guide 2 Telescope bench.

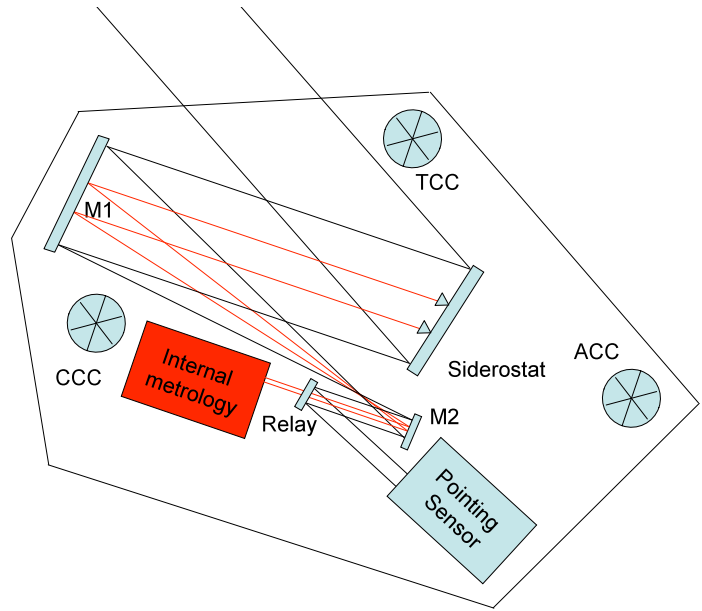

Fig. 5. Guide 2 Telescope layout.

As the attitude of SIM-Lite changes in inertial space, the fine stage of the siderostat mechanism tracks the Guide 2 star. This technique allows us to keep the Guide 2 star locked on the pointing camera, at the intersection of 4 pixels within a few milli-arcseconds, while measuring the larger dynamics of the Spacecraft attitude change (about 1 arcsecond) with the metrology sensor. The latter sensor has a much more linear response than the camera would have, if we did not keep the star at the cross-hair of 4 pixels. Both the CCD based pointing sensor and the metrology system tracking the angular position of the siderostat have accuracy close to 20 micro-arc-seconds over short time period [10].

## 2.5. Metrology Truss

The SIM-Lite metrology truss has 5 primary nodes:
- two double fiducials (DCC) imbedded into the Science interferometer collecting mirrors define the Science baseline,
- two triple fiducials (TCC) in front of the Guide 1 interferometer collecting mirrors define the Guide baseline,
- one single fiducial, called the apex corner-cube (ACC), out of the Science/Guide baseline plane is used to measure any out of plane motion of the interferometric baselines.

One of the Guide 1 TCC fiducial and the apex corner-cube are attached to the Guide 2 telescope bench, such that the nominal line of sight of the Guide 2 telescope is aligned with the metrology truss link between this Guide 1 fiducial and the apex corner-cube. A third fiducial (CCC) on the Guide 2 Telescope is used to monitor the bench motion in the less sensitive degrees of freedom.

## 2.6. Overall spacecraft configuration

The SIM-Lite spacecraft is designed around a 7-meter-long truss-based precision support structure (PSS). Sub-assemblies interface to the various nodes at the intersection of the PSS struts. The configuration is shown in Fig. 6.

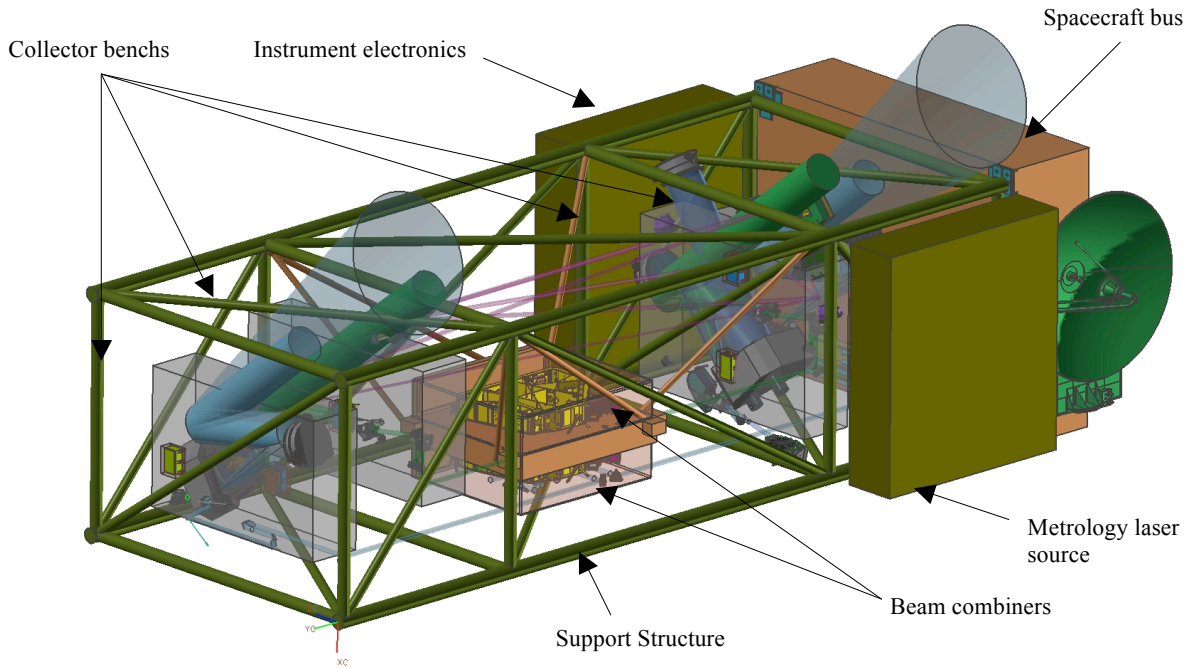

Fig. 6. Conceptual configuration of SIM-Lite.

The first subsystem is the spacecraft bus that provides power, telecommunication and avionics for maneuvering the entire spacecraft. The spacecraft bus is mechanically isolated from the PSS through torsion beams to reduce the jitter induced by the reaction wheels. The next subsystem is the instrument bus that contains the metrology source, the instrument computer and electronics.

The following subsystem is the first Science interferometer collector bench. This precision graphite bench carries the Siderostat with the fiducial that defines the interferometer baseline, a 30 cm beam compressor and the optical delay line that is used to equalize the internal delay in the Science interferometer.

Next is the Guide collector bench that hosts both the first set of collecting optics for the Guide 1 interferometer and the Guide 2 Telescope. Similarly to the Science collecting optics, the Guide 1 collecting optics consist of a fiducial and a 30 cm beam compressor, and the Guide 2 telescope consists of a siderostat for initial acquisition of the Guide 2 star and a 30 cm beam compressor with a focal plane camera.

The next set of assemblies is the two astrometric beam combiners (ABC) [11], stacked on top of each other, near the center of the PSS. Each ABC recombines the light from the two collectors, forms images of the stars on the angle tracking camera for real-time pointing control and interferences fringes onto the fringe tracking camera for real-time pathlength control as well as astrometric measurement. The ABC also hosts the internal metrology system that monitors the optical path from the ABC combiner to the collecting fiducials.

The last two subsystems are the second Guide collecting bench and Science collecting bench, similar to the other two collecting benches.

## 3. ASTROMETRIC PERFORMANCE

The 5 year mission will be divided in a few major programs. Table 1 shows a possible mission allocation, based on similar allocation from SIM PQ:

Table 1. Mission time allocation.

| Task | Targets | Mission |
|---|---|---|
| Tier 1 (1Earth) | 65 stars | 46% |
| Tier 2 | 1050 stars | 5% |
| Young Stars | 67 stars | 2% |
| Grid | 44,000 tiles | 9% |
| Quasars | 50 quasars | 1.5% |
| Wide Angle | 8,300 hours | 19% |
| S/C Slewing | 63,500 slews | 14.5% |
| Alignment/Cal | 45 min/day | 3% |

### 3.1. Deep planetary survey

The deep planetary survey (also known as Tier-1 survey) will focus on relatively few (less than one hundred) nearby stars of the main sequence (F, G, K and M types), within 10 parsecs from the Sun. The main objective is to identify planetary system with Earth-like planets in the habitable zone around these Sun-like stars.

This deep survey requires the highest possible astrometric accuracy, below the single micro-arc-second. This accuracy is achieved by multiple visits to the target stars. SIM-Lite will achieve 1.4 µas differential error between Tier 1 target position and a set of Reference stars [12] per 12-chop visit. Most Tier-1 target will require more than 12 chops per visits, the average being around 60 chops per visits to lower the instrument accuracy down to the required level for detection of 1-Earth mass planet. A 60 chop visit to a magnitude 6 Tier 1 Target would require less than two hours of observation time. The target would then be observed several hundred times during the lifetime of the mission.

Fig. 7 illustrates the sequence used for the Tier-1 observation. The sequence starts with 20 seconds of observation time on the target star T, during which interference fringes are being collected. The observation is followed by about 15 seconds to slew and reposition the two Science siderostats and the optical delay line to acquire fringes on the first Reference star R1. After 40 seconds of observation on R1, the interferometer is pointed back to the same target star T to be re-observed for 20 seconds. Then, the interferometer is pointed to the second Reference star R2, we observe for 40 seconds, and point back to the target star. We continue re-pointing and observing between the target stars and the other Reference stars R3 and R4. Finally, we repeat the sequence from the beginning. During the entire sequence, the Guide 1 interferometer and the Guide 2 telescope are locked on their respective stars, monitoring changes of the instrument attitude in inertia space.

Fig. 7. Narrow Angle observation sequence.

In order to determine accurately the orbit parameters of the planets, 200 visits per Tier 1 target star will be scheduled over the 5 year mission, 100 visits per baseline orientation.

The following criteria are used on the selection of the Tier 1 targets to maximize the instrument performance:
- First, the Tier 1 Target star magnitude shall be brighter than 6. Fig. 8 shows a histogram of 65 candidate Tier 1 targets, all of candidates meet the magnitude 6 criteria.
- Second, at least three Reference stars of magnitude lower than 9 can be found in a 1.25 degree radius field around the Tier 1 target star. Fig. 8 shows a histogram of the mean Reference star flux for 28 Reference star groups. The effective magnitude on the Reference stars is lower than 9 for all but 3 of Tier 1 target.

Fig. 8. Left: Star Visible magnitude for the first 65 best Tier-1 targets. Right mean Reference star flux for 28 set of Reference stars

The timing breakdown for a typical 12 T-R-T chop visit to a Tier-1 target is:
- 360 seconds of retargeting time between stars (24 times 15 seconds).
- 240 seconds of total integration time on the Tier 1 target per visit (12 times 20 seconds).
- 480 seconds of total integration time on the set of Reference stars (12 times 20 seconds).

The total scheduled time per 12-chop visit to a Tier 1 target is 1,080 seconds and increases to 5,400s for a 60 chop visit. Assuming that 46% of the mission time be allocated to the deep planetary survey, the number of planetary systems that can be surveyed would be:

$$46\% * 5y * 365d * 86,400s / 5,400s / 200\text{visits} \sim 67 \text{ planetary systems} \qquad (3)$$

### 3.2. Broad planetary survey

The Broad planetary survey (also known as Tier-2 survey) will focus on a larger (more than one thousand) set of stars of many types (including O, B, A, F, binary) to cover the diversity of planetary systems [2]. The main objective is to increase our knowledge on the nature and the evolution of planetary systems in their full variety.

This broad survey can be achieved with a reduced astrometric accuracy of five micro-arc-seconds. This accuracy can be achieved by short visits to the target stars. SIM-Lite will achieve an instantaneous 5uas differential error between Tier 2 target position and a set of Reference stars with a single 60 second visit.

In order to determine accurately the orbit parameters of the planets, 100 visits per Tier 2 target star will be scheduled over the 5 year mission, 50 visits per orientation.

The following criteria are used on the selection of the Tier 2 targets to maximize the instrument performance:
- First, the Tier 2 Target star magnitude shall be brighter than 10.5, in order to limit the observation time to 60 seconds per visit.
- Second, to maximize the mission efficiency, the Tier 2 target needs to be located near an existing Tier-1 target star from the deep survey or a young planetary system. By doing so, no additional mission time is required to point the space-craft toward the Tier-2 target and the same set of Reference stars can be used. Surveys of the stars within 2 degrees from the existing 65 Tier-1 targets, indicate that more than 1000 stars are closer than 200 parsecs and brighter than magnitude 10.5.

The timing breakdown for a visit to a magnitude 10.5 Tier-2 target is:
- 15 seconds retargeting time to point at the Tier-2 star.
- 60 seconds of integration time on the Tier-2 target.

The total scheduled time per visit to a Tier 2 target is 75 seconds. Assuming that 5% of the mission time be allocated to the broad planetary survey, the number of planetary systems that can be surveyed would be:

$$5\% * 5y * 365d * 86,400s / 75s / 100\text{visit} \sim 1050 \text{ planetary systems} \qquad (4)$$

### 3.3. Young planetary systems

The young planetary system survey (also known as Young star survey) will focus on relatively few (less than one hundred) nearby solar type stars with ages below 100 millions years, within 100 parsecs from the Sun. The main objective is to understand the frequency of Jupiter-mass planets and the early dynamical evolution of planetary systems.

This young planetary system survey can be achieved with a reduced astrometric accuracy of four micro-arc-seconds. SIM-Lite will achieve the 4uas differential error between young star position and a set of Reference stars by 4-chops for every visit. A 4 chop visit to a magnitude 11 young star requires about eight minutes of observation time.

In order to determine accurately the orbit parameters of the planets, 100 visits per young star will be scheduled over the 5 year mission, 50 visit per orientation.

The following criteria are used on the selection of the young planetary system targets to maximize the instrument performance:
- First, the young star magnitude shall be brighter than 11. Fig. 9 shows the number of candidate young stars versus magnitude. About 80 candidates meet the brighter than magnitude 11 constraint.

- Second, at least one Guide star of magnitude brighter than 7 exist in a 1 degree radius field around the young star. Most of the young stars being fainter than 7th magnitude, are not good candidate for tracking the attitude of the Guide baseline.
- Third, at least three Reference stars of magnitude lower than 9 can be found in a 1.25 degree radius field around the young star. This magnitude constraint on the Reference stars is met for all the Tier 1 target and will likely be met for all the young stars too.

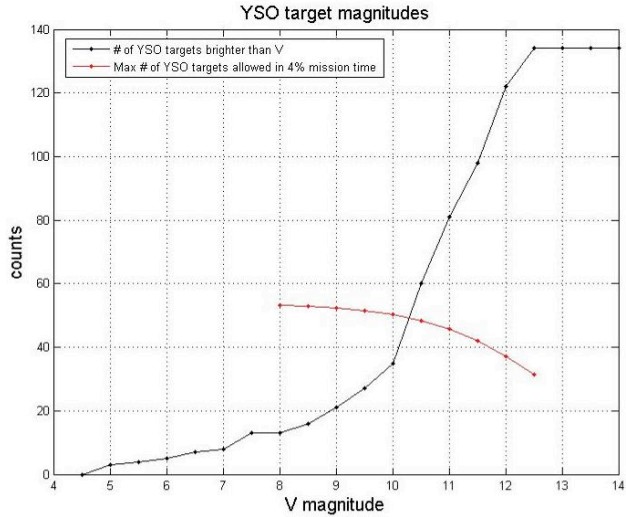

Fig. 9. How many targets can be in the Young Star Survey?

The timing breakdown for a 4 T-R-T chop visit to a young planetary sytem is:
- 120 seconds of total retargeting time between stars.
- 250 seconds of total integration time on the young star.
- 100 seconds of total integration time on the set of Reference stars.

The total scheduled time for a visit to a young star is 470 seconds. Assuming that 2% of the mission time be allocated to the young planetary system survey, the number of planetary systems that can be surveyed would be:

$$2\% * 5y * 365d * 86{,}400s / 470s / 100\text{visit} \sim 67 \text{ planetary systems} \quad (5)$$

### 3.4. Global astrometric grid

Most of the non-planetary astrophysics programs that SIM-Lite will execute rely on an absolute reference frame called the grid. SIM-Lite will build an all-sky astrometric grid made of about 1300 pre-selected stars and about 50 quasars, to an end of mission accuracy of 4 micro-arc-seconds in positions at the mean epoch and in parallaxes.

The full field of regard of SIM-Lite is 15 degrees in diameter, covering about 6 Grid stars in average at any position of the sky. SIM-Lite will observe the Grid stars sequentially, going through the 1300 tiles centered on the 1300 Grid stars. During the 5 years, SIM-Lite will go 34 times through the set of 1300 tiles. Since during each tile, an average of 6 Grid stars will be observed, at the end of 5 year period, each Grid star will have been observed 34*6 ~ 200 times.

The set of Grid stars already identified for SIM PlanetQuest is fully appropriate for SIM-Lite. The key criteria for the selection were that the Grid stars be K giant stars, at least 500 parsecs away, with no stellar companions, spread quasi-uniformly over the sky to maximize the grid stability and performance. The set of 1300 Grid stars has a median magnitude of V = 9.9. One additional criteria is the existence of a Guide star of magnitude brighter than 7 within in a couple of degrees from each Grid stars. The majority of the Grid stars have a magnitude in the V = 9 – 10.5 range, therefore they are not good candidate for tracking the attitude of the Guide baseline.

Each visit to a typical tile will consist of a single G1–G2–G3…G6–G1 chop through the 6 Grid stars. The time required for a visit to a tile of magnitude 10.5 Grid stars is:
- 210 seconds of total retargeting time between 6 Grid stars.
- 105 seconds of total integration time on the set of Grid stars.

The total scheduled time for a visit to a tile is 315 seconds. The grid build-up will be completed after 34 visits to the 1300 distinct tiles over the 5 year mission. One could estimate the fraction of the mission time that will be required to complete the 44,000 tiles:

$$34 \text{visits} * 1{,}300 \text{tiles} * 315s / 5y * 365d * 86{,}400s \sim 9\% \qquad (6)$$

The grid will be anchored to an absolute inertial reference frame made of a set of distant quasars. Observing the 50 quasars will require an additional 1.5% of the mission time.

### 3.5. Wide angle astrometry

The Wide Angle astrometry program covers a wide range of topics in Galactic and extragalactic astronomy, such as formation and dynamics of our Galaxy, calibration of the cosmic distance scale, and fundamental stellar astrophysics. The global astrometric performance will be limited to the 4 micro-arc-seconds end of mission grid error for bright objects up to magnitude 14. For dimmer, the performance would vary depending on the observation time allocated to the object, reaching 20 micro-arc-second astrometric performance on magnitude 20 objects after about 50 hours of integration.

In order to compensate for parallax effect, an average of 100 visits per object will be scheduled over the 5 year mission, 50 visits per orientations. The integration time on the object for each visit will vary with magnitude and expected performance, all the way to two hours on the dimmest objects. There is no hard limitation on the magnitude, but integration time for object fainter than magnitude 20 require hours just for initial acquisition of the fringes.

SIM-Lite like SIM PlanetQuest would allocate 19% of the 5 year mission to Wide Angle astrometry. The corresponding 8,300 hours would be allocated based on Table 2.

Table 2. Total observation time (hours) versus measurement accuracy.

| Magnitude | 10 | 12 | 14 | 16 | 18 | 19 | 20 |
|---|---|---|---|---|---|---|---|
| 5 µas | 0.6 | 1.0 | 3.0 | 15.4 | | | |
| 6 µas | 0.3 | 0.5 | 1.6 | 8.1 | 66.8 | | |
| 8 µas | | 0.3 | 0.7 | 3.7 | 30.2 | 102.0 | |
| 10 µas | | | 0.5 | 2.2 | 17.7 | 59.8 | 230.1 |
| 15 µas | | | 0.3 | 1.2 | 7.4 | 24.6 | 94.4 |
| 20 µas | | | | 0.9 | 5.0 | 13.9 | 51.7 |
| 30 µas | | | | 0.6 | 3.4 | 8.7 | 24.8 |

### 3.6. Non-observational time

The total number of spacecraft maneuvers required to re-point the instrument to new star fields to complete the various observation programs is shown in Table 3. The average slew angles will be 7 degrees. It will take 6 minutes to slew the spacecraft by angles of 7 degrees in average and to settle the instrument attitude at the end of the maneuver.

Table 3. Total required number of spacecraft maneuvers to complete the SIM-Lite mission.

| Task | Targets | Visits | S/C Slews |
|---|---|---|---|
| Tier 1 (1Earth) | 65 stars | 200 | 13,000 |
| Young Stars | 65 stars | 100 | 6,500 |
| Grid and Wide Angle | 1,300 tiles | 34 | 44,000 |
| Total | | | 63,500 |

The estimation of required number of spacecraft maneuvers assumes that the observation of the Planetary Broad Survey Tier 2 targets, the Quasars and Wide Angle Astrometry targets do not require additional dedicated slews. The total mission time to be allocated for spacecraft maneuvers is 14.5% of the mission:

$$63,500 \text{slews} * 360s / (5y * 365d * 86,400s) \sim 14.5\% \tag{7}$$

## 4. CONCLUSION

SIM-Lite is a great alternative to SIM-PQ which, while producing most of the original science, greatly reduces the cost. This will make SIM-Lite a much more affordable mission in which NASA can invest in the next few years. During the next two years, we will finish the Brassboard development of all major components, at which point, we will be ready to start building the SIM-Lite flight hardware.

## ACKNOWLEDGEMENTS

The research described in this publication was performed at the Jet Propulsion Laboratory of the California Institute of Technology, under contract with the National Aeronautics and Space Administration. The authors would like to thank the SIM instrument team for their contribution to the SIM-Lite mission concept.